\def\<{\langle}
\def\>{\rangle}

\newcommand{\ra}{\;\raise1.0pt\hbox{$'$}\hskip-6pt\partial\;}
\newcommand{\lo}{\;\overline{\raise1.0pt\hbox{$'$}\hskip-6pt\partial}\;}

\newcommand{\Abs}[1]{\begin{abstract} #1 \end{abstract}}

\newcommand{\mktt}{  }

\documentclass[twocolumn, times, twocolappendix]{aastex631}

\usepackage{graphicx, epsfig, natbib, color, times, bm, amsmath,
hyperref, multirow, harpoon, amssymb, mathtools, mathrsfs, xcolor,
academicons, booktabs, soul}

\usepackage[ruled,linesnumbered]{algorithm2e}

\begin{document}

\title{ Accelerating spherical harmonic transforms for a large number of
sky maps}

\author{Chi Tian}
\affiliation{School of Physics and Optoelectronics Engineering, Anhui
University, 111 Jiulong Road, Hefei, Anhui, China 230601.}

\author{Siyu Li}
\affiliation{Key Laboratory of Particle and Astrophysics, Institute of
High Energy Physics, CAS, 19B YuQuan Road, Beijing, China, 100049.}

\author{Hao Liu$^*$}
\affiliation{School of Physics and Optoelectronics Engineering, Anhui
University, 111 Jiulong Road, Hefei, Anhui, China 230601.}
\affiliation{Key Laboratory of Particle and Astrophysics, Institute of
High Energy Physics, CAS, 19B YuQuan Road, Beijing, China, 100049.}

\email{$^*$Corresponding author, ustc\_liuhao@163.com}

\Abs{

The spherical harmonic transform is a powerful tool in the analysis of
spherical data sets, such as the cosmic microwave background data. In
this work, we present a new scheme for the spherical harmonic transforms
that supports both CPU and GPU computations, which is especially
efficient on a large number of sky maps. By comparing our implementation
with the standard \textsl{Libsharp-HEALPix} program, we demonstrate 2-10
times speedup for the CPU implementation, and up to 30 times speedup
when a state-of-the-art GPU is employed. This new scheme's software
package is available via an open source GitHub repository.

}
\mktt

\section{Introduction}\label{sec:intro}

The spherical harmonic transform (SHT) decomposes signals on a sphere
into sets of spherical harmonic coefficients based on the spherical
harmonic functions. As an analog to the Fourier transform, the SHT
extracts scale-dependent features on the sphere and has been extensively
used in analyzing spherical data in various areas of science, including
cosmology and the cosmic microwave background (CMB) science. For
example, by performing SHTs on the CMB anisotropy data and fitting with
the concordance cosmological model ($\Lambda$CDM), we are able to make
precise estimations about the cosmological parameters. The pipeline for
processing CMB polarization data with SHT is also crucial for detecting
the primordial gravitational waves. Recent developments in cosmological
surveys in CMB temperature and polarization
anisotropies~\citep{2003SPIE.4843..284K, 2006astro.ph..4069T,
2010ASSP...14..127R, 2010SPIE.7741E..1SN, 2011arXiv1110.2101K,
2012SPIE.8452E..1EA, 2012SPIE.8442E..19H, 2016arXiv161002743A,
doi.10.1093.nsr.nwy019, 2019JCAP...02..056A} pose new challenges to data
analysis, which often requires processing a large amount of CMB mock
data. Therefore, it is important to reduce the time cost of SHTs on
numerous spherical maps.

The optimizations for the SHT have been intensively studied in the past
years. Optimized codes, such as \textsl{Libsharp}
\citep{2013A&A...554A.112R} and \textsl{SHTns}
\citep{2013GGG....14..751S} have been used in various cosmological
studies. However, it is difficult to further accelerate the SHT on a
single sky map, because the most time-consuming part of the SHT is
effectively a BLAS (basic linear algebra subprograms) level 2 problem,
which is not very suitable for high-performance computing (HPC). In this
work, we discover that when there are many input maps, the core part of
SHT can be updated from BLAS level 2 to level 3, which is not only
appropriate for HPC but also has many established solutions. Following
this idea, we propose a highly efficient SHT algorithm that is able to
process a large number of spherical maps in a batch. We will demonstrate
the power of this new scheme by carrying out SHTs on mock CMB
temperature and polarization maps with our open source software package
{---}
\textsl{fastSHT}\footnote{\url{https://github.com/liuhao-cn/fastSHT},
and the version corresponding to this work is~\cite{Liuhao-cn2022-vz}.},
whose performance will be compared with the SHT toolkit
\textsl{Libsharp}, which is the SHT engine of \textsl{HEALPix}
\citep{2005ApJ...622..759G} and \textsl{Healpy}, where the latter
provides a python interface.

Also note that the recent developments in the Graphic Processing Units
(GPUs) pave the way to further accelerating the traditional spherical
data analysis pipeline, which was mostly based on computations with
CPUs. Pioneering studies such as \cite{10.1007/978-3-642-29737-3_40}
have demonstrated the power of GPUs in computing the SHT on a single
map. In this work, we will explore both the GPU and CPU implementations
of our new SHT algorithm with the state-of-the-art computer hardware.

This paper is organized as follows, we first introduce the mathematical
basis of the SHT in section~\ref{sec:basis}, and then use CMB
temperature and polarization maps to explain in detail how to accelerate
SHTs for a large number of spherical maps. The software implementation
of the new SHT scheme is presented in section~\ref{sec:performance},
where comprehensive memory usage estimations and performance benchmarks
with CPUs and GPUs are included. Then a new iteration scheme based on
our software implementation is introduced in section~\ref{sec: new
iter}, which significantly improve the quality of iterative estimation.
Finally, we conclude and discuss in section~\ref{sec:discussion}.

\section{Mathematical basis}
\label{sec:basis}

The basic algorithm of fast SHT has been described in a number of
literatures~\citep{SHT..Mohlenkamp1999, SHT..Suda..02,
2005IJMPD..14..275D, SHT..Huffenberger_2010, SHT..2011A&A...526A.108R,
SHT..S2HAT, SHT..2013ITIP...22.2275M}. In particular, some tips of
acceleration are summarized in~\cite{2013GGG....14..751S}. Till the
present day, there seems to be no significant changes to these basic
aspects of the algorithm, particularly for the single-map computation;
thus, for convenience of reading, the basis of the algorithm is briefly
summarized below, followed by our new batch SHT scheme.

\subsection{SHT of CMB temperature maps}
\label{sub:temperature}

The forward SHT decomposes a scalar field on a sphere, such as the CMB
temperature map $T(\theta,\phi)$, into spherical harmonic coefficients
$a_{\ell m}$s, following the equation
\begin{align}
a_{\ell m} =& \sum_{\theta\phi}T(\theta,\phi)Y_{\ell m}(\theta, \phi)d\sigma,
\end{align}
where $\theta$ and $\phi$ are discretized polar and azimuthal angle
respectively, and $d\sigma=4\pi/N_{\rm pix}$ is the area of one pixel,
which is a constant in the HEALPix pixelization scheme.

Considering a series of maps denoted by index $k$, and decomposing
$Y_{\ell m}(\theta, \phi)$ into the associate Legendre polynomials
$P_{\ell}^m [\cos(\theta)]$ with a phase term\footnote{Here the
associated Legendre polynomials are usually re-normalized to the
spherical harmonics as $|P_{\ell}^m [\cos(\theta)]| = |Y_{\ell
m}(\theta, \phi)|$, and this kind of re-normalized $P_{\ell}^m$ are also
referred to as $\lambda_{\ell}^m$ and shortened as ``lam'' in several
source codes, e.g., HEALPix.}, the above equation becomes
\begin{align}
\label{eq:mm}
a_{\ell m}^k =& \sum_{\theta\phi}T^k(\theta,\phi)P_{\ell}^m[\cos(\theta)]
e^{-im\phi}d\sigma 
\\ \nonumber
=& \sum_{\theta\phi}T^k(\theta,\phi)P_{\ell}^m[\cos(\theta)]
\left[e^{-i\frac{2\pi mn}{N}} e^{-im\phi_0(\theta)}\right]d\sigma 
\\ \nonumber
=& \sum_{\theta}P_{\ell}^m[\cos(\theta)]
\left[\sum_{n}T^k(\theta,n)e^{-i\frac{2\pi mn}{N}}\right] 
e^{-im\phi_0(\theta)}d\sigma \\ \nonumber
=& \sum_{\theta}P_{\ell}^m[\cos(\theta)]\mathcal{T}_m^k(\theta),
\end{align}
where $N$ is the number of pixels in the iso-latitude ring located at
$\theta$, $n$ is the local index of the pixels inside the ring, and
$\phi_0(\theta)$ is the $\phi$-coordinate of the first pixel in each
ring. The value $\mathcal{T}_m^k(\theta) \equiv
\left[\sum_{n}T^k(\theta,n) e^{-i\frac{2\pi mn}{N}} \right]
e^{-im\phi_0(\theta)}$ is the combination of Fourier transforms of
$T(\theta, \phi)$ with constant phase shifts, which can be computed by a
forward fast Fourier transform (FFT). The result of FFT is phase-shifted
as above and properly rearranged into matrix forms, which is called the
``FFT mapping'' in the following part of this work. The FFT mapping
usually takes about 10-15\% of the total computation time, whereas the
most time-consuming part is the last row of the above equation. If there
is only one input map, the operation in the last row effectively involve
only matrix-to-vector multiplications, since the subscript $m$ does not
participate in the computation. This kind of task is called a BLAS level
2 problem. Unfortunately, it is well known that a BLAS level 2 problem
cannot be significantly accelerated, whereas a BLAS level 3 problem,
such as a matrix-to-matrix multiplication (MM), can be effectively
optimized by fully exploiting the cache-based architecture (see
\cite{10.1145/1377603.1377607}).

When there are a large number of spherical maps, the last row of
eq.~\eqref{eq:mm} becomes a matrix-to-matrix multiplication that
contracts the $\theta$-dimension, which looks like
\begin{eqnarray} \label{equ:mm_forward}
\left(a_{\ell}^k\right)_m = \left(P_{\ell}^\theta \cdot
\mathcal{T}_{\theta}^k\right)_m.
\end{eqnarray}
$m$ is written outside the brackets to indicate that it is not part of
the matrix multiplication. This reformulation suggests that matrix
multiplications could be employed to accelerate SHTs of a batch of sky
maps.

Similarly, the backward transform is given by equations below:
\begin{align}\label{equ:backward equ}
T^k(\theta,\phi) =& \sum_{\ell m}a_{\ell m}^kP_{\ell}^m[\cos(\theta)]
e^{im\phi} \\ \nonumber
=& \sum_{\ell m}a_{\ell m}^kP_{\ell}^m[\cos(\theta)]
e^{i\frac{2\pi mn}{N}}e^{im\phi_0(\theta)} \\ \nonumber
=& \sum_{m=0}^{l_{\textrm{max}}}e^{i\frac{2\pi mn}{N}} \left[e^{im\phi_0(\theta)}
\sum_{l=m}^{l_{\textrm{max}}}a_{\ell m}^kP_{\ell}^m[\cos(\theta)] \right] 
\\ \nonumber
=& \sum_{m=0}^{l_{\textrm{max}}}\mathcal{T}_{m}^k(\theta)e^{i\frac{2\pi mn}{N}} ,
\end{align}
where $\mathcal{T}_{m}^k(\theta)=\left[e^{im\phi_0(\theta)}
\sum_{l=m}^{l_{\textrm{max}}}a_{\ell m}^kP_{\ell}^m[\cos(\theta)]
\right]$ should be computed by another matrix multiplication that
contracts $\ell$, which looks like
\begin{eqnarray}\label{equ:mm backward}
\left(\mathcal{T}^k_\theta\right)_m= \left(a_{\ell}^k \cdot P^{\ell}_\theta\right)_m,
\end{eqnarray}
and then the last row of eq.~(\ref{equ:backward equ}) is computed by a
backward-FFT ring by ring, with $\mathcal{T}_{m}^k(\theta)$ being the
input FFT-coefficients.

To summarize, for a batch of input sky maps, the forward SHT can be
accomplished by computing the FFT, FFT-mapping, and MMs; and the
backward SHT can be computed in reverse order by performing the MM,
FFT-mapping, and FFT. Compared to the regular scheme that processes the
maps one by one, this newly designed pipeline fully exploits the
optimizations from matrix multiplications and can therefore be more
efficient. Moreover, this new scheme also allows more effective usage of
the GPU and enables a much bigger room for acceleration.

\subsection{Simplified matrix form}
\label{sub:matrix form}

Eq.~\eqref{equ:mm_forward} and eq.~\eqref{equ:mm backward} are the cores
of our new SHT method, and they can be further simplified by defining
the following matrices:
\begin{align}\label{equ:mat form def}
\bm{A}_m = (a^k_\ell)_m,\,\,
\bm{\mathcal{T}}_m = (\mathcal{T}^k_i)_m,\,\,
\bm{P}_m = (P^i_\ell)_m.
\end{align}
Note that these matrices are usually non-square. The spherical harmonic
transforms are then given by the following matrix multiplications at
each $m$, with the convention that the superscript of the first matrix
is contracted with the subscript of the second matrix.
\begin{align}\label{equ:matrix form}
\bm{A}_m&= \bm{P}_m\bm{\mathcal{T}}_m
\\ \nonumber
\bm{\mathcal{T}}_m&= \bm{P}^t_m\bm{A}_{m}.
\end{align}
Then $\bm{\mathcal{T}}_m$ is connected to the temperature map by the
FFT.

\subsection{SHT of the CMB polarization maps}
\label{sub:pol case}

In the context of rotations on the sphere, the $Q$- and $U$-Stokes
parameters can be decomposed into spin $\pm2$ spherical
harmonics~\citep{PhysRevD.55.1830, 0004-637X-503-1-1} as
\begin{equation} \label{Q_lm+iU_lm}
    Q(\mathbf{\hat{n}})\pm i U(\mathbf{\hat{n}}) = 
    \sum_{l,m} a_{\pm2,lm}\;{}_{\pm2}Y_{lm}(\mathbf{\hat n}),
\end{equation}
where $_{\pm2}Y_{lm}(\mathbf{\hat n})$ are the spin $\pm2$ spherical
harmonics. Similar to eq.~(\ref{equ:mat form def}--\ref{equ:matrix
form}), the spin $\pm2$ spherical harmonic coefficients $a_{\pm2,lm}$
are given by the following matrix multiplications:
\begin{align}
\bm{A}_{\pm2,m} &= \bm{F}_{\pm2,m}(\bm{\mathcal{Q}}_m \pm i\,\bm{\mathcal{U}}_m),
\end{align}
where $\bm{\mathcal{Q}}_m$ and $\bm{\mathcal{U}}_m$ are the phase
shifted Fourier transforms of the $Q$ and $U$ Stokes parameters of the
$k$-th map at the $i$-th ring, similar to $\bm{\mathcal{T}}_m$; and
$\bm{F}_{\pm2,m}$ are the spin$\pm2$ spherical harmonics without the
phase term, so $\bm{F}_{\pm2,m}$ are real value matrices.

Because the $E$- and $B$-mode spherical harmonic coefficients are
defined as
\begin{align} \label{equ:alm-eb} 
a_{E,lm} &= -(a_{2,lm} + a_{-2,lm})/2,
\nonumber \\ 
a_{B,lm} &= i(a_{2,lm} - a_{-2,lm})/2,
\end{align}
we get {\scriptsize
\begin{align}\label{equ:aeab complex}
\bm{A}^E_m &= -\frac{1}{2}\left[ \bm{F}_{+2,m}\bm{\mathcal{Q}}_m + 
\bm{F}_{+2,m}(i\,\bm{\mathcal{U}}_m) + \bm{F}_{-2,m}\bm{\mathcal{Q}}_m - 
\bm{F}_{-2,m}(i\,\bm{\mathcal{U}}_m) \right]
\\ \nonumber
& = -\frac{1}{2}\left[(\bm{F}_{+2,m}+\bm{F}_{-2,m})\bm{\mathcal{Q}}_m + 
(\bm{F}_{+2,m}-\bm{F}_{-2,m})(i\,\bm{\mathcal{U}}_m)\right]
\\ \nonumber
\bm{A}^B_m &= \frac{i}{2}\left[ \bm{F}_{+2,m}\bm{\mathcal{Q}}_m + 
\bm{F}_{+2,m}(i\,\bm{\mathcal{U}}_m) - \bm{F}_{-2,m}\bm{\mathcal{Q}}_m + 
\bm{F}_{-2,m}(i\,\bm{\mathcal{U}}_m) \right]
\\ \nonumber
& = \frac{i}{2}\left[(\bm{F}_{+2,m}-\bm{F}_{-2,m})\bm{\mathcal{Q}}_m + 
(\bm{F}_{+2,m}+\bm{F}_{-2,m})(i\bm{\mathcal{U}}_m)\right].
\end{align}
}

By defining
\begin{align}
\bm{F}_{+,m} &= -(\bm{F}_{+2,m}+\bm{F}_{-2,m})/2
\\ \nonumber
\bm{F}_{-,m} &= -(\bm{F}_{+2,m}-\bm{F}_{-2,m})/2,
\end{align}
eq.~(\ref{equ:aeab complex}) can be simplified to:
\begin{align}
\bm{A}^E_m &= \bm{F}_{+,m}\bm{\mathcal{Q}}_m + 
\bm{F}_{-,m} (i\,\bm{\mathcal{U}}_m)
\\ \nonumber
i\,\bm{A}^B_m &= \bm{F}_{-,m}\bm{\mathcal{Q}}_m + 
\bm{F}_{+,m} (i\,\bm{\mathcal{U}}_m),
\end{align}
which can be further simplified to the following matrix form:
\begin{align}\label{equ:aeab mat form}
\begin{pmatrix}
\bm{A}^E \\
i\bm{A}^B
\end{pmatrix}_m
=
\begin{pmatrix}
\bm{F}_+ & \bm{F}_-
\\
\bm{F}_- & \bm{F}_+
\end{pmatrix}_m
\begin{pmatrix}
\bm{\mathcal{Q}} \\
i\,\bm{\mathcal{U}}
\end{pmatrix}_m.
\end{align}
From eq.~(\ref{equ:aeab mat form}), we immediately get the backward
transform as:
\begin{align}\label{equ:aeab mat form inv}
\begin{pmatrix}
\bm{\mathcal{Q}} \\
i\,\bm{\mathcal{U}}
\end{pmatrix}_m
=
\begin{pmatrix}
\bm{F}_+ & \bm{F}_-
\\
\bm{F}_- & \bm{F}_+
\end{pmatrix}^t_m
\begin{pmatrix}
\bm{A}^E \\
i\bm{A}^B
\end{pmatrix}_m.
\end{align}
Therefore, $i\bm{A}^B_m$ and $i\,\bm{\mathcal{U}}_m$ are more natural
than $\bm{A}^B_m$ and $\bm{\mathcal{U}}_m$ in the spherical harmonic
transforms of polarized data.

From eqs.~(\ref{equ:aeab mat form}--\ref{equ:aeab mat form inv}), it is
also convenient to derive pure pixel-domain decompositions of the $E$-
and $B$-modes:
\begin{align}\label{equ:aeab mat form pixel}
\begin{pmatrix}
\bm{\mathcal{Q}}\\
i\,\bm{\mathcal{U}}
\end{pmatrix}_m^{E,B}
&=
\begin{pmatrix}
\bm{F}_+ & \bm{F}_-
\\
\bm{F}_- & \bm{F}_+
\end{pmatrix}^t_m
\bm{\lambda}^{E,B}
\begin{pmatrix}
\bm{A}^E \\
i\bm{A}^B
\end{pmatrix}_m
\\ \nonumber
&=
\begin{pmatrix}
\bm{F}_+ & \bm{F}_-
\\
\bm{F}_- & \bm{F}_+
\end{pmatrix}^t_m
\bm{\lambda}^{E,B}
\begin{pmatrix}
\bm{F}_+ & \bm{F}_-
\\
\bm{F}_- & \bm{F}_+
\end{pmatrix}_m
\begin{pmatrix}
\bm{\mathcal{Q}}\\
i\,\bm{\mathcal{U}}
\end{pmatrix}_m,
\end{align}
where superscripts $E$, $B$ means to keep only the $E$- and $B$-mode
components, and $\bm{\lambda}^{E,B}$ is a diagonal matrix with either 0
or 1 along the diagonal line. For $\bm{\lambda}^{E}$, only the first
half of the diagonal line is equal to 1, and for $\bm{\lambda}^{B}$,
only the second half of the diagonal line is equal to 1.

The above form might be the most elegant representation of the essence
of $E$- and $B$-modes, because it shows that, the only difference
between the $E$- and $B$-modes is to divide the identity matrix $\bm{I}$
into two parts and keep either the first or the second part non-zero.

Apparently, eqs.~(\ref{equ:aeab mat form}--\ref{equ:aeab mat form
pixel}) are block form representations of larger matrices, and the
multiplication of each pair of elements means another matrix
multiplication. The combined larger matrices can be written as
\begin{align}
\bm{\mathcal{S}}_m = 
\begin{pmatrix}
\bm{\mathcal{Q}} \\
i\,\bm{\mathcal{U}}
\end{pmatrix}_m
,\;\;
\bm{\mathcal{F}}_m = \begin{pmatrix}
\bm{F}_+ & \bm{F}_-
\\
\bm{F}_- & \bm{F}_+
\end{pmatrix}_m,\;\;
\bm{\mathcal{A}}_m = 
\begin{pmatrix}
\bm{A}^E \\
i\bm{A}^B
\end{pmatrix}_m,
\end{align}
then the SHTs of the polarized data can be simply computed by the
following matrix multiplications:
\begin{align}
\bm{\mathcal{A}}_m &= \bm{\mathcal{F}}_m \bm{\mathcal{S}}_m
\\ \nonumber
\bm{\mathcal{S}}_m &= \bm{\mathcal{F}}_m^t \bm{\mathcal{A}}_m,
\end{align} 
where $\bm{\mathcal{F}}_m$ is a real matrix but might be non-square.
When an EB-decomposition is required, it can be done as
\begin{align}
\bm{\mathcal{S}}_m^{E,B} &= (\bm{\mathcal{F}}_m^t \bm{\lambda}^{E,B} \bm{\mathcal{F}}_m)
\bm{\mathcal{S}}_m,
\end{align}
where $\bm{\mathcal{F}}_m\bm{\mathcal{F}}_m^t$ and $\bm{\lambda}^{E,B}$
are both real, diagonal and contain only 0 or 1. These equations are
enough to handle the spherical harmonic transforms of polarized data in
either pixel or harmonic domains.

\section{Code implementation and performance} 
\label{sec:performance}

In this section, we will introduce and benchmark our code,
\textsl{fastSHT}, which is an optimized implementation of the new SHT
scheme introduced above. Its core part is written in FORTRAN for both
the CPU and GPU implementations, and the FORTRAN subroutines are wrapped
by the software \textsl{f90wrap} \citep{Kermode2020-f90wrap} to expose
Python interfaces to users. The GPU feature can be optionally turned
on/off by the user when compiling the code. For convenience, we will
from now on refer to the name \textsl{fastSHT-CPU} as the code complied
only with CPU; and the name \textsl{fastSHT-GPU} as the code compiled
with the GPU feature enabled. The \textsl{fastSHT-CPU} code employs the
Intel MKL library for the FFT and general matrix multiplication (gemm)
to carry out the operations mentioned above; whereas
\textsl{fastSHT-GPU} benefits not only from the faster GPU matrix
multiplications, but also from the more efficient FFT mappings and
iterative operations. The internal loops of \textsl{fastSHT-GPU} are
parallelized using OpenACC (Open Accelerators), significantly reducing
the coding complexity.

\subsection{Implementation detail}
\label{sub:imple}
\begin{algorithm}
\caption{fastSHT algorithm -- non-iterative}\label{alg1}
\KwData{Input sky maps as matrix: $\bm{S}$} 
\KwResult{Spherical harmonic coefficients: $a_{\ell m}^k$}
$\bm{F}^k_{\theta m}=\rm{Batch~FFT}$$(\bm{S})$

\For{$m = 0$ \KwTo $\ell_{\rm max}$}{

\For{$\theta = 0$ \KwTo $N_{\rm rings}$}{

\For{$\ell = 0$ \KwTo $\ell_{\rm max}$}{

compute Legendre polynomials $P_m(l,\theta)$  
}
\For{$k = 0$ \KwTo $N_{\rm maps}$}{

remap $\bm{F}^k_{\theta m}$ to $\mathcal{T}_m(\theta, k)$
}
}

$a_m(k, \ell) = \mathrm{gemm}(P_m(\ell,\theta), \mathcal{T}_m(\theta,k))$

}

\end{algorithm}

Our \textsl{fastSHT} code is based on the \textsl{HEALPix} pixelization
scheme, where the sphere is divided into iso-latitude rings, and then
into equal-area pixels. The iso-latitude rings in this pixelization
scheme naturally discretize the polar angle $\theta$. We give the pseudo
code of our new SHT algorithm in Algorithm~\ref{alg1}. The definitions
of the auxiliary fields ($P_m$ and $\theta_m$) are in section
\ref{sub:temperature}, and the variable $N_{\rm rings}$ is the total
number of iso-latitude rings. In addition, to accelerate SHT as much as
possible for a large number of sky maps, the following points should be
taken into account in the implementation:

\begin{enumerate}
\item All SHT operations are done with real numbers to maximize
efficiency.

\item \label{itm:fft1} The FFT needs to be performed in batch, and any
unnecessary memory operations, i.e, read/write operations during the
FFT, should be done by the FFT program itself. This means that the FFT
implementation should support I/O distances and I/O strides.

\item \label{itm:fft2} In a FFT implementation like FFTW or Intel MKL, a
plan/handle will be created before computing, in which the FFT size is
fixed to maximize the speed. This plan/handle should be reused for rings
of the same size until all rings of the same size are done.

\item Since the FFTs are done ring by ring for all maps, in the GPU
implementation, once an FFT on one ring is done, the memory copy from
the output of the FFT to the device memory can be done asynchronously
with the unfinished FFT operations, which are performed on the CPU only.
This overlaps the memory copy with the FFT operations and saves time
significantly.

\item \label{itm:fft3} By default, the primary dimension\footnote{The
``primary dimension'' is the dimension with continuous storage. For
example, in FORTRAN this is the first dimension, and in C this is the
last dimension.} of the FFT result is occupied by $m$. However, because
$m$ does not participate in matrix multiplication, and current ``gemm''
programs requires continuous primary dimension; a transpose procedure is
needed, which should be done by the FFT program via setting of the FFT
I/O strides.

\item The most important point is that, the FFT-to-$a_{\ell m}$
operations (eqs.~(\ref{equ:mm_forward} \& \ref{equ:mm backward})) should
be done by the matrix multiplication, which is a part of lower level
BLAS libraries. These libraries are the core parts of numerous HPC
applications, and therefore are usually fully supported and highly
optimized. BLAS implementations, such as the Intel MKL library, the
cuBLAS BLAS library, and third party libraries like Eigen, OpenBLAS and
clBLAS, have been widely used. Moreover, these implementations also
allow various deployments, from laptop to super computers, from Windows
to Linux, and from CPU platforms to GPU platforms. In this study, we use
the ``dgemm'' subroutine from the Intel MKL BLAS library for the CPU
code and the ``cublasDgemm'' subroutine from the NVIDIA cuBLAS library
for the GPU code. These subroutines and their sisters\footnote{In the
naming system of BLAS, the first letter is one of s, d, c or z, which
mean single/double precision real numbers and single/double precision
complex numbers, respectively, and ``gemm'' means general matrix
multiplication.} have been proven to be the top-notch BLAS libraries.

\item \label{itm:8} To fully exploit the power of the ``dgemm'' or
``cublasDgemm'' implementation, the storage scheme should be carefully
designed. We employ a novel storage scheme of $a_{\ell m}$s, which
compactifies complex data into real matrices (see appendix~\ref{app:tech
tips} for details). Moreover, the built-in read-write support of the
``dgemm'' should be used as much as possible to reduce redundant
read-write operations. In other words, the reading, conversion,
computing, writing, and scaling are done by the ``dgemm'' as much as
possible, and different types of the SHTs are implemented by changing
the constants and the order of calling the ``dgemm'' subroutine.

\item When running with some iterative schemes, e.g., by setting a
non-zero $N_{\textrm{iter}}$ parameter in \textsl{HEALPix} , it is
important to restrict the iterations in between
$\mathcal{T}_{m\theta}^k$ and $a_{\ell m}^k$ (eq.~\eqref{equ:mm_forward}
and eq.~\eqref{equ:mm backward}), because the forward and backward FFT
transforms between $\mathcal{T}_{m}^k(\theta)$ and $T^k(\theta,\phi)$
are lossless. Meanwhile, the iterative updating of $a_{\ell m}$s should
be done by the ``dgemm'' via a optimal setting of parameters to save
time.

\item The program should support forward and backward transforms of
$a_{\ell m}^E$ and $a_{\ell m}^B$ separately. In some data processing
pipelines, each step may only need one of
them~\citep{2018arXiv181104691L, 2019arXiv190400451L,
Liu_2019_EB_general}, so the speed of SHTs can be increased by 100\% for
free.

\item The ring optimization~\citep{2013GGG....14..751S} should be used,
which means for a given $m$, the rings with extremely low amplitudes of
$P_{\ell m}$ are ignored. This saves about $15\%$ of time, which is
consistent with~\citep{2013GGG....14..751S}. Note that the ring
optimization affects not only $P_{\ell m}$, but also the FFT-mapping.

\item When $N_{\textrm{side}}>512$, it is necessary to scale some
$P_{\ell m}$s during recursions, because the recursions start from the
diagonal elements ($\ell=m$) that contain factor $\sin^{m}(\theta)$,
which can cause arithmetic underflow for a 64-bit floating number at
high $m$. However, a normal scaling method causes considerable
performance loss when $P_{\ell m}$ is required to be a matrix.
Therefore, the following dedicated methods should be adopted: for
$N_{\textrm{side}}\le 512$, we do not use the scaling; and for higher
resolutions, the recursion starts from ``safe positions'' that allow the
scaling to be ignored. Here the ``safe positions'' mean the
$\ell$-positions from which the amplitudes of $P_{\ell m}[\cos(\theta)]$
are larger than a given threshold, such as $10^{-20}$. These positions
are computed once and saved as an input table
$\ell_{\textrm{safe}}(m,\theta)$.

\item For the polarized SHT that deals with both E and B-modes
(QU-to-EB), there are two possible orders of computation: 1) firstly
QU-to-E and then QU-to-B; 2) firstly Q-to-EB and then U-to-EB.  The
second one is adopted because it helps to reduce the number of
FFT-mapping operations by 50\% and saves the memory compared to the
first one.

\end{enumerate}

\subsection{Memory cost}

In Table \ref{tab:mem cost}, we show the peak memory cost of performing
SHTs on $1000$ maps with $N_{\rm side} = 128$ for both the CPU and GPU
implementations.  Input maps, output spherical harmonic coefficients
($a_{\ell m}$s), and all necessary buffers are included in the
estimation.  Note that the memory cost approximately scales
quadratically with $N_{\rm side}$, and linearly with the number of maps;
thus, Table \ref{tab:mem cost} provides a baseline for a quick
estimation of the peak memory cost for other cases. Generally speaking,
to achieve the best performance, the memory cost of the workflow should
be close to the physical memory capacity, especially for
\textsl{fastSHT-GPU}.

\begin{table}[!ht]
    \centering
    \begin{tabular}{|c|c|r|r|}
    \hline
    Polarization    & Iteration  &    CPU memory & GPU memory \\ \hline
    No              & No         &  4.4 (4.9) GB & 3.2 GB     \\ \hline
    No              & Yes        &  5.9 (4.9) GB & 4.7 GB     \\ \hline
    Yes             & No         &  8.8 (7.2) GB & 5.9 GB     \\ \hline
    Yes             & Yes        & 11.8 (7.2) GB & 8.9 GB     \\ \hline
    \end{tabular}
    \caption{ The peak memory cost of performing SHTs on $1000$ maps
    with $N_{\rm side} = 128$. Input maps, output spherical harmonic
    coefficients ($a_{\ell m}$s), and all necessary buffers are included
    in the estimation. The numbers in brackets are the CPU memory costs
    of the GPU implementation. Due to allocating pinned memory for
    buffer arrays, \textsl{fastSHT-GPU} consumes slightly more CPU
    memory in row 1. }
    \label{tab:mem cost}
\end{table}

\subsection{Performance}\label{sub:performance}

We test the performance of the \textsl{fastSHT} code on a docker
virtualized machine with state-of-the-art hardware --- it contains an
Intel Platinum 8336C CPU and an NVIDIA Tesla A100 GPU. We are limited to
using 24 CPU cores since the performance of the CPU version is also
limited by other hardware subsystems.\footnote{ For both
\textsl{fastSHT} and \textsl{Healpy}, there is typically a CPU core
number threshold after which the performance stops improving. Because
this threshold is determined by both the hardware and software
environments, it should be discovered by evaluating each computation
system separately. To assist the readers with such tests, we have
included a script called "run\_nproc.sh" in the GitHub repository. } The
compilers used in these tests are the Intel FORTRAN compiler and the
NVIDIA PGI (Portland Group Inc.) FORTRAN compiler for the CPU and GPU
versions, respectively. Even though GPUs are known to be more powerful
at single-precision floating-point (FP32) computations than
double-precision ones (FP64), we stick to computing with double
precision in all our tests to fully meet the precision requirement of
various cosmological simulations and observations. The versions of
compilers and software used in this article to generate the benchmarks
are: Intel One API (2022.0.2), NVIDIA HPC SDK (22.3), Healpy (15.2 with
default installation), fastSHT (\cite{Liuhao-cn2022-vz} with default
installation).

In our tests, we notice that the \textsl{fastSHT-GPU} usually introduces
non-negligible initialization overhead. This overhead includes the
initialization of the cuBLAS library and the pre-allocation of cache
arrays. Moreover, when a GPU is involved, the CPU buffer arrays are
allocated in the form of pinned arrays, which ensures faster
communications with the GPU memory at the cost of a larger allocation
overhead. Fortunately, the \textsl{fastSHT-GPU} only needs to be
initialized once for any amount of computation in similar batches. For
example, an extremely large number of maps can be packed into several
smaller batches, and the \textsl{fastSHT-GPU} initialization overhead
will only happen once for the first batch, and then all follow-up
batches are free from this overhead. This feature makes this
initialization overhead distinct from other in-computation costs, such
as the data transfer cost between the CPU and GPU. On the other hand,
for a complicated analysis that involves massive floating point
calculation, the initialization overhead is relatively negligible (see
Table~\ref{tab:fix-eb test} for a realistic example). Therefore, in all
of the performance evaluations presented in this work, we choose to
separate the time costs of the initialization overhead and main SHT
computations, and exclude this overhead when quantifying the
acceleration factor for the \textsl{fastSHT-GPU} code.

\begin{figure}[!ht]
  \centering
  \includegraphics[width=0.48\textwidth]{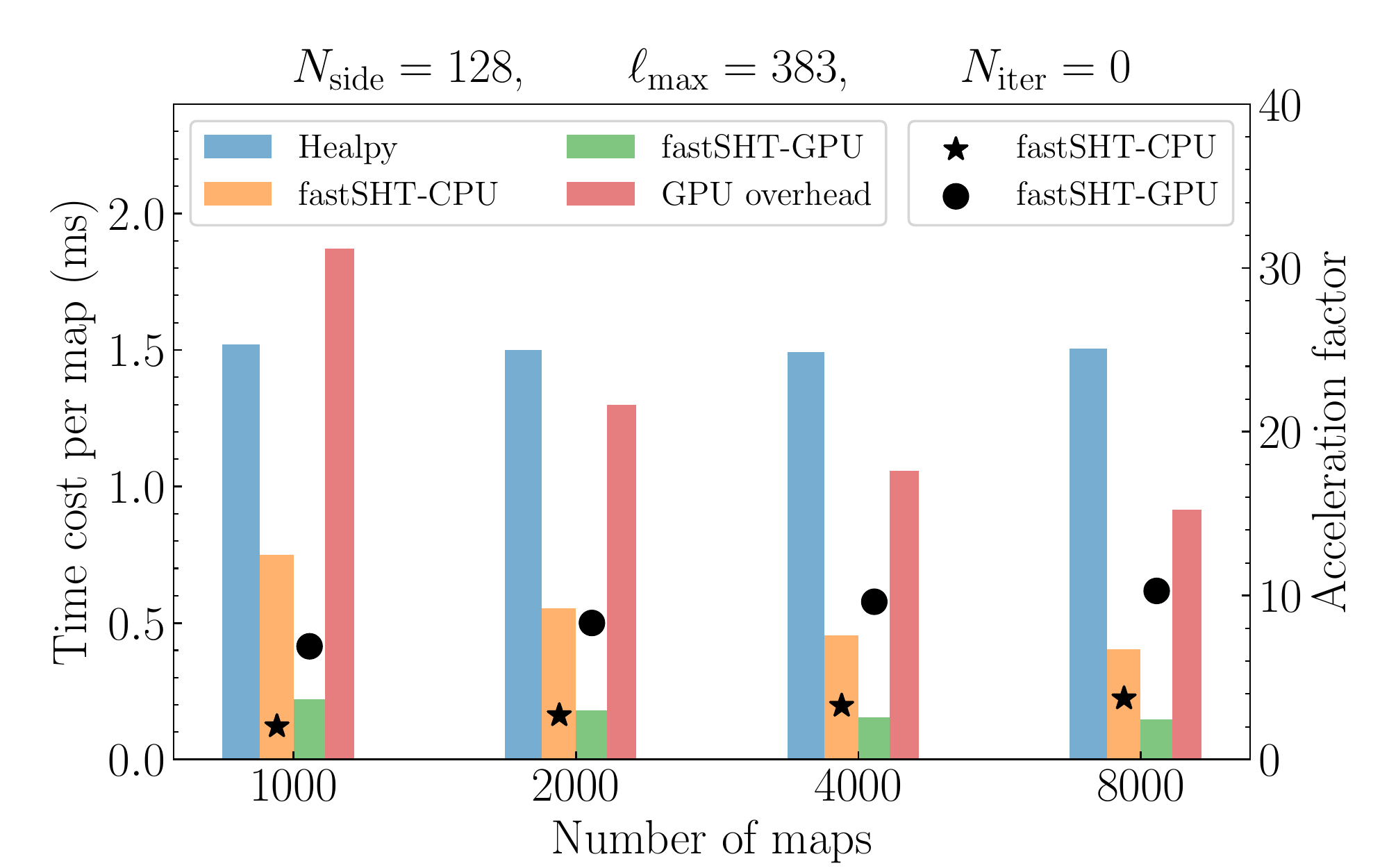}
  
  \includegraphics[width=0.48\textwidth]{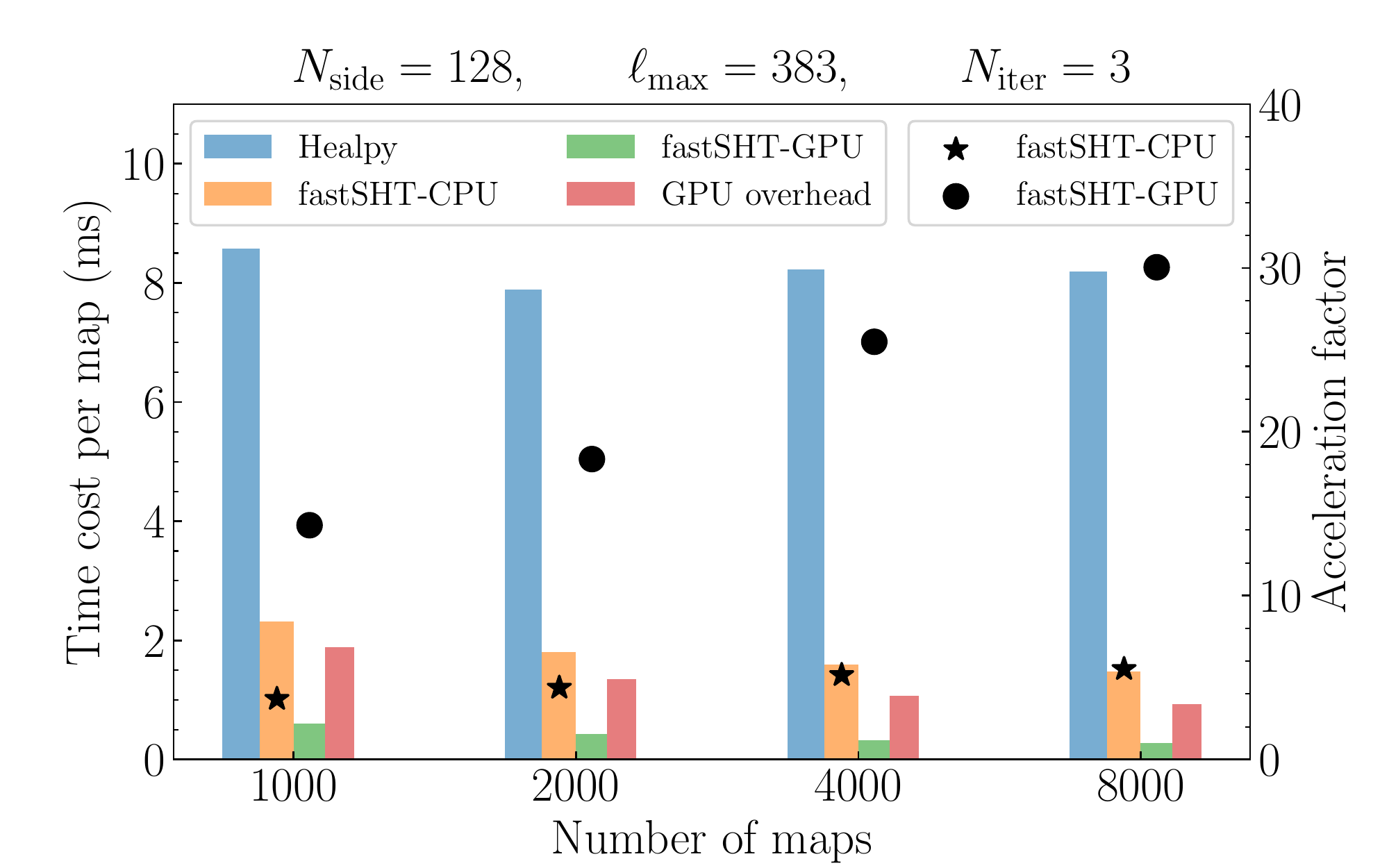}
  \caption{ The non-polarized SHT performance tests for \textsl{Healpy},
  \textsl{fastSHT-CPU} and \textsl{fastSHT-GPU} with $N_{\rm side} =
  128$, $\ell_{\rm max} = 383$ and $N_{\rm maps}$ varies from 1000 to
  8000. The upper/lower panels show the case without/with iterations.
  The colored bars (belong to the left $y$-axis) give the time costs per
  map, and the stars/dots (belong to the right $y$-axis) show the
  acceleration factors relative to \textsl{Healpy}. The CPU computations
  are done with 24 cores.  }
  \label{fig:t2alm}
\end{figure}

In Figure~\ref{fig:t2alm}, we compare the time costs of \textsl{fastSHT}
with \textsl{Healpy} for the non-polarized cases with/without
iterations. The input map resolution is $N_{\rm side} = 128$ and
$\ell_{\rm max} = 3N_{\rm side}-1$. As explained above, the
\textsl{fastSHT-GPU} time cost includes the CPU-GPU memory exchange
time, but excludes the initialization overhead. In the upper panel of
Figure~\ref{fig:t2alm}, we focus on the case without iteration:
\textsl{fastSHT-CPU} gives acceleration factors between 2.0$\times$ and
3.7$\times$, increasing with $N_{\rm maps}$; whereas
\textsl{fastSHT-GPU} gives acceleration factors close to 10$\times$.
When a standard iteration scheme is performed to both \textsl{fastSHT}
and \textsl{Healpy} to improve the SHT accuracy, the advantage of
\textsl{fastSHT} is more manifest because the initialization overhead
happens only once, and the time cost of memory copying is also
relatively smaller, as shown in the lower panel of
Figure~\ref{fig:t2alm}. An acceleration factor bigger than 5.0$\times$
can be achieved for \textsl{fastSHT-CPU} with more than $4000$ maps,
whereas the acceleration factor of \textsl{fastSHT-GPU} reaches about
30$ \times$ at $8000$ maps.

\begin{figure}[!htb]
  \centering
  \includegraphics[width=0.48\textwidth]{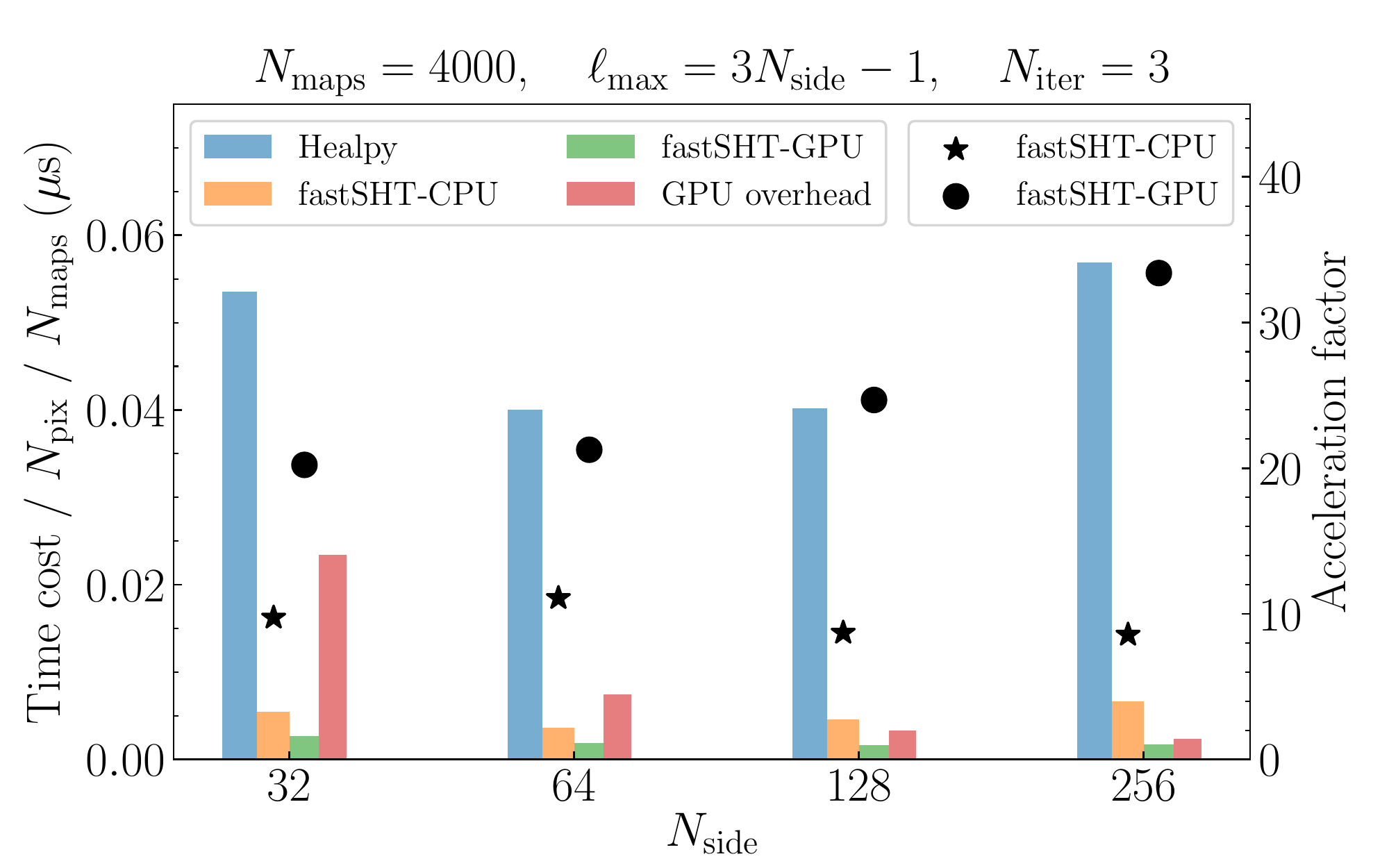}
  
  \includegraphics[width=0.48\textwidth]{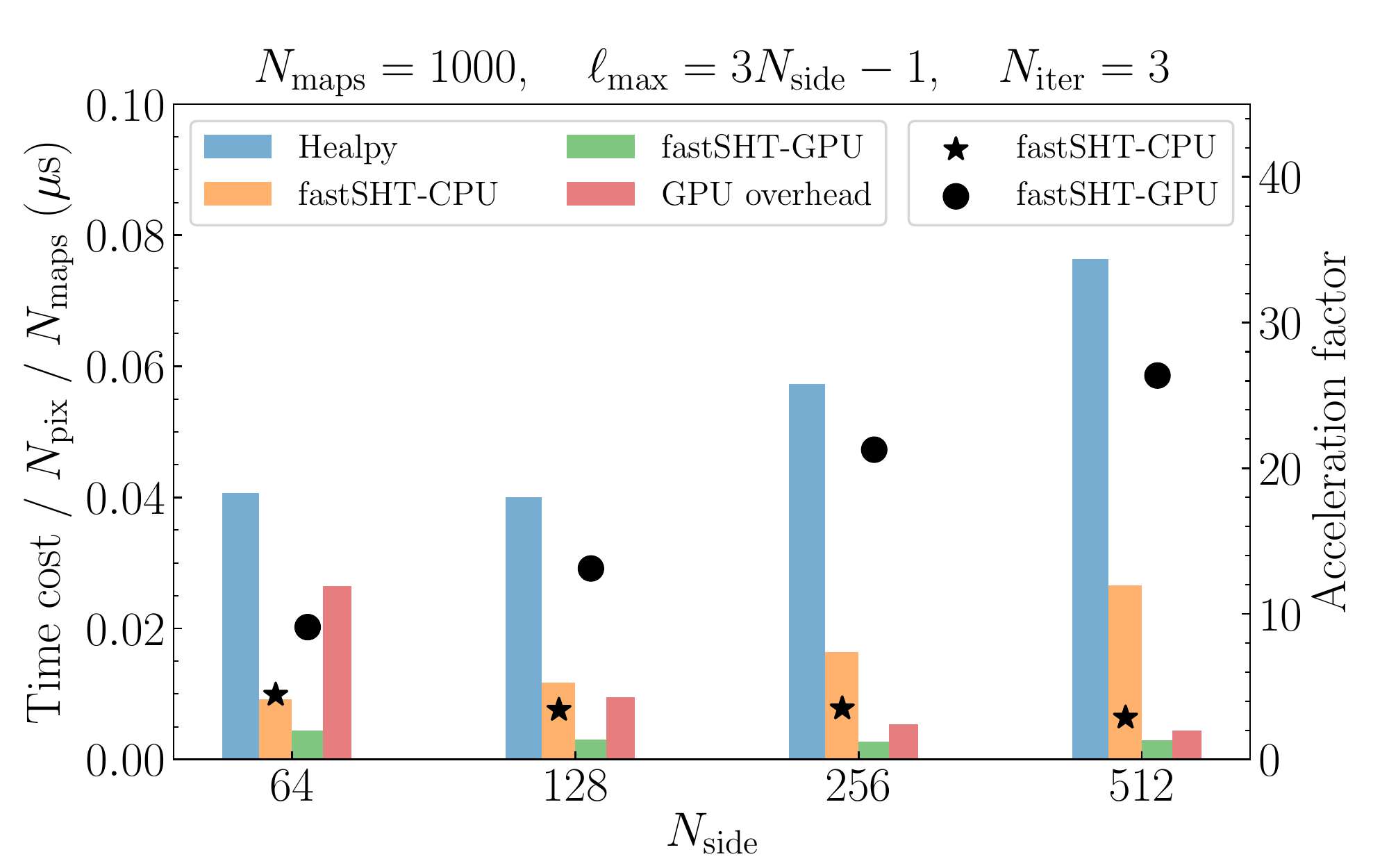}
  
  \includegraphics[width=0.48\textwidth]{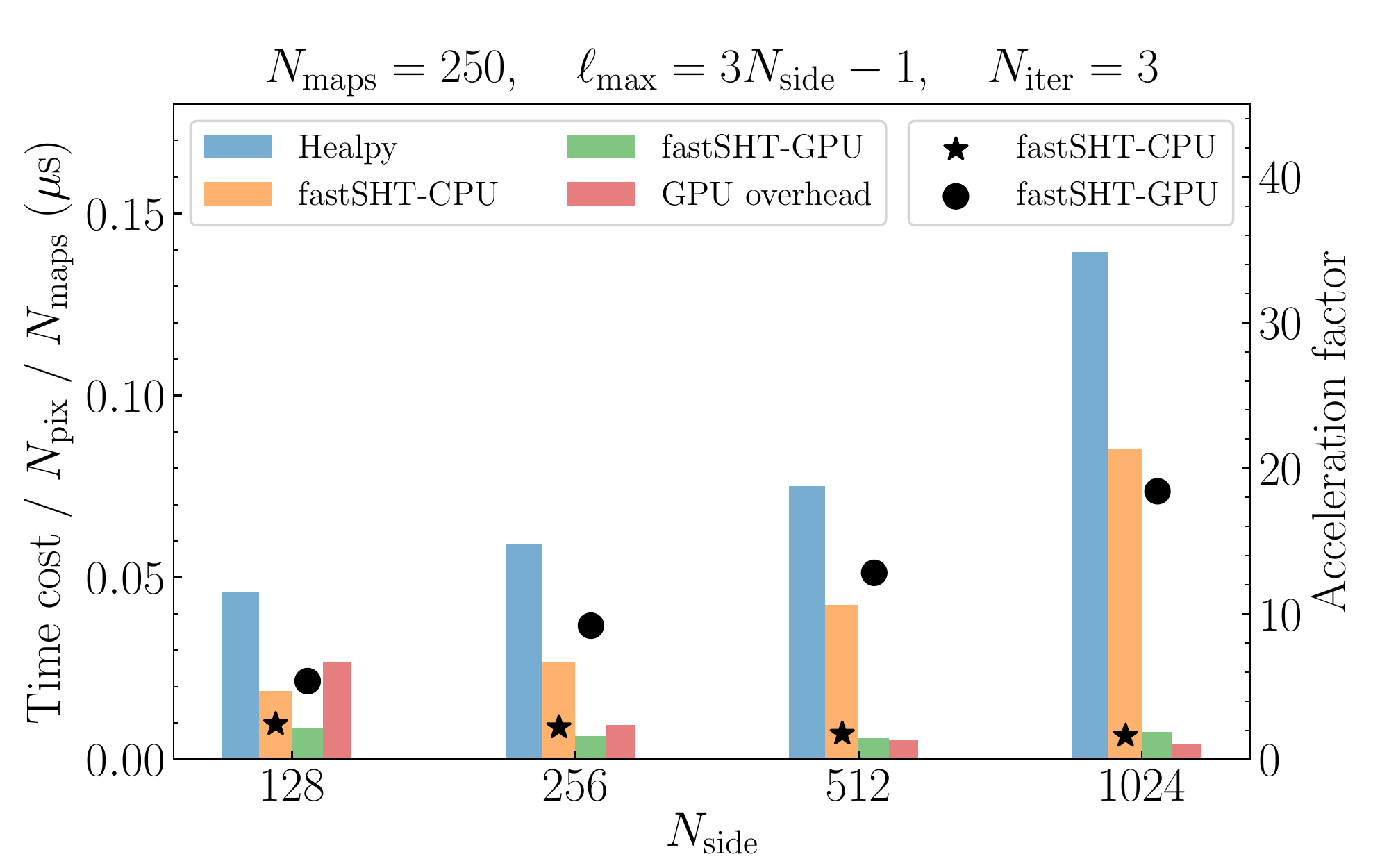}
  \caption{ The SHT performance tests without polarization for
  \textsl{Healpy}, \textsl{fastSHT-CPU} and \textsl{fastSHT-GPU}, with
  $N_{\rm side}$ varies from 32 to 1024, $\ell_{\rm max} = 3 N_{\rm
  side}  - 1$, and $N_{\rm iter}=3$. The numbers of maps are 4000 $\sim$
  250 (top to bottom) respectively, which are determined by the GPU
  memory limit at the highest resolution in each panel. The colored bars
  (belong to the left $y$-axis) give the time costs per sky map per
  pixel, and the stars/dots (belong to the right $y$-axis) show
  \textsl{fastSHT-CPU} and \textsl{fastSHT-GPU}'s acceleration factor
  relative to \textsl{Healpy}, respectively. }
  \label{fig:t2alm_ns}
\end{figure}

In Figure~\ref{fig:t2alm_ns}, we show how the performance of
\textsl{fastSHT} changes with the map resolution $N_{\rm side}$. The
panels from top to bottom show the results of $N_{\rm maps} = 4000,
1000, 250$ respectively. The number of maps is different for different
panels because we are limited by the GPU memory capacity. Due to better
parallel efficiencies of \textsl{Healpy} at higher resolutions, the
acceleration factor of \textsl{fastSHT-CPU} is gradually suppressed at
higher resolutions (but still $>$1.6 in all cases). The
\textsl{fastSHT-GPU}, on the contrary, exhibits an increasing
acceleration factor with $N_{\rm side}$, which is up to 34$\times$ at
higher resolutions. The exact performance improvement remains to be
checked by a future hardware with larger GPU memory capacity. However,
the tendency is evident: according to Figures~\ref{fig:t2alm} --
\ref{fig:t2alm_ns}, more GPU memory usage leads to higher performance;
thus, using \textsl{fastSHT-GPU} is recommended for larger number of
maps; meanwhile, if the number of map is fixed, then a higher resolution
leads to a higher performance, until one hits the GPU memory limit.

In particular, to demonstrate potential applications of our new SHT
scheme in a realistic situation, we perform a pipeline of fixing the
EB-leakages by the best blind estimator \citep{2018arXiv181104691L,
Liu_2019_EB_general}, which not only includes all types of SHTs
(forward, backward, temperature, polarization, iterative scheme), but
also involves a linear regression procedure and a lot of memory updating
operations. We choose to work with $3$ iterations and at two different
resolutions: the first one is $N_{\rm side} = 256$ and $N_{\rm maps} =
2000$, and the second one is $N_{\rm side} = 512$ and $N_{\rm maps} =
500$. The number of CPU cores is still 24, and the time costs are listed
in Table~\ref{tab:fix-eb test}, which shows CPU acceleration factors of
4.5$\times$ and 3.2$\times$ and GPU acceleration factors of 13$\times$
and 14$\times$, respectively for the two tests. It is worth noting that
the GPU initialization overheads in these two tests are less than
$5\,s$, accounting for only 5--7$\%$ of the total time. This is a direct
consequence of the massive floating-point computation required by the
pipeline, and the \textsl{fastSHT-GPU} outperforms the traditional
scheme significantly even after including the overhead, suggesting a
great application potential of our SHT scheme in a computationally
intensive CMB data processing.
\begin{table}[!ht]
    \centering
    \begin{tabular}{cccccc}
    $N_{\rm side}$ & $N_{\rm maps}$ & $t_1 $ & $t_2$ & $t_3$ & Overhead \\ \hline
    $256$ & $2000$ & $815\, s$ & $181\,s$ ($4.5\times$) & $62\,s$ ($13\times$) & $4.8\,s$ \\ \hline
    $512$ &  $500$ & $1325\,s$ & $416\, s$ ($3.2\times$) & $95\,s$ ($14\times$) & $4.9\,s$ \\ \hline
    \end{tabular}
    \caption{The time costs for \textsl{Healpy} ($t_1$),
    \textsl{fastSHT-CPU} ($t_2$), and \textsl{fastSHT-GPU} ($t_3$) in a
    pipeline that fixes the EB-leakage. The CPU part is done with 24
    cores. The GPU overhead is presented in the last column, and the
    acceleration factors relative to \textsl{Healpy} are given in
    brackets. }
    \label{tab:fix-eb test}
\end{table}

Lastly, we notice that \textsl{Libsharp} also supports an optimization
option when processing many input maps based on a straightforward idea:
compute the $P_{\ell m}$s only once, and then reuse them for each input
map. Nevertheless, the calculation of the $P_{\ell m}$s only takes a
small fraction of the total computational time in both \textsl{Libsharp}
and \textsl{fastSHT}, and indeed we see no significant improvement when
enabling this feature in \textsl{Libsharp} in our tests. Meanwhile, this
is an advanced feature and is not included in \textsl{Healpy}. As a
result, we ignore this feature in the \textsl{Libsharp} and do not cache
$P_{\ell m}$s in our \textsl{fastSHT} code.

\section{An improved iteration scheme}\label{sec: new iter}

The SHT accuracy can be improved by an iterative solution. In this work,
two iteration schemes are considered: one is the ``traditional'' mode
adopted by the \textsl{HEALPix} , in which the $a_{\ell m}$s are updated
outside the $m$-loop; and the other one is the ``immediate'' mode, in
which the $a_{\ell m}$s are updated immediately within the $m$-loop. The
pseudo codes with more details for these two schemes are described in
Algorithm \ref{alg2} and \ref{alg3} respectively.

\begin{algorithm}
\caption{Fast SHT algorithm for many sky maps with the traditional
iterative method} \label{alg2}
\KwData{Input maps as matrix: $\bm{S}$ 
\newline Number of iterations: $N_{iter}$}
\KwResult{Spherical harmonic coeficients: $a_{\ell m}$}

$\bm{F}^k_{\theta m}=\mathrm{Batch~FFT}(\bm{S})$

\For{$n= 1$ \KwTo $N_{iter}$}{

\For{$m= 0$ \KwTo $\ell_{\rm max}$}{

\For{$\theta= 0$ \KwTo $N_{\rm rings}$}{

\For{$\ell= 0$ \KwTo $\ell_{\rm max}$}{
 compute Legendre polynomials $P_m(\ell,\theta)$  
}

}

$\delta (\mathcal{T}^k_\theta)_m = (\mathcal{T}^k_\theta)_m - (a^k_\ell \cdot P^\ell_\theta)_m$

update $\bm{F}^k_{\theta m}$ from $\delta (\mathcal{T}^k_\theta)_m$

}

\For{$m= 0$ \KwTo $\ell_{\rm max}$}{

\For{$\theta\leftarrow 0$ \KwTo $N_{\rm rings}$}{
\For{$\ell= 0$ \KwTo $\ell_{\rm max}$}{
compute Legendre polynomials $P_m(\ell,\theta)$  
}

\For{$k= 0$ \KwTo $N_{\rm maps}$}{

remap $\bm{F}^k_{\theta m}$ to $(\mathcal{T}_{\theta}^k)_m$
}
}
$\left(a_{\ell}^k\right)_m = \left(P_{\ell}^\theta\cdot \mathcal{T}_{\theta}^k\right)_m$
}
}
\end{algorithm}

\begin{algorithm}
\caption{Fast SHT algorithm for many sky maps with the immediate
iterative method} \label{alg3}
\KwData{Input maps as matrix: $\bm{S}$ 
\newline Number of iterations: $N_{iter}$}
\KwResult{Spherical harmonic coeficients: $a_{\ell m}$}

$\bm{F}^k_{\theta m}=\mathrm{Batch~FFT}(\bm{S})$

\For{$n= 1$ \KwTo $N_{iter}$}{

\For{$m= 0$ \KwTo $\ell_{\rm max}$}{

\For{$\theta= 0$ \KwTo $N_{\rm rings}$}{

\For{$\ell= 0$ \KwTo $\ell_{\rm max}$}{
 compute Legendre polynomials $P_m(\ell,\theta)$  
}

}

$\delta (\mathcal{T}^k_\theta)_m = (\mathcal{T}^k_\theta)_m - (a^k_\ell \cdot P^\ell_\theta)_m$

update $\bm{F}^k_{\theta m}$ from $\delta (\mathcal{T}^k_\theta)_m$

\For{$k= 0$ \KwTo $N_{\rm maps}$}{

remap $\bm{F}^k_{\theta m}$ to $(\mathcal{T}_{\theta}^k)_m$
}

$\left(a_{\ell}^k\right)_m = \left(P_{\ell}^\theta\cdot \mathcal{T}_{\theta}^k\right)_m$
}

}
\end{algorithm}

\begin{figure}[!htb]
  \centering
  \includegraphics[width=0.23\textwidth]{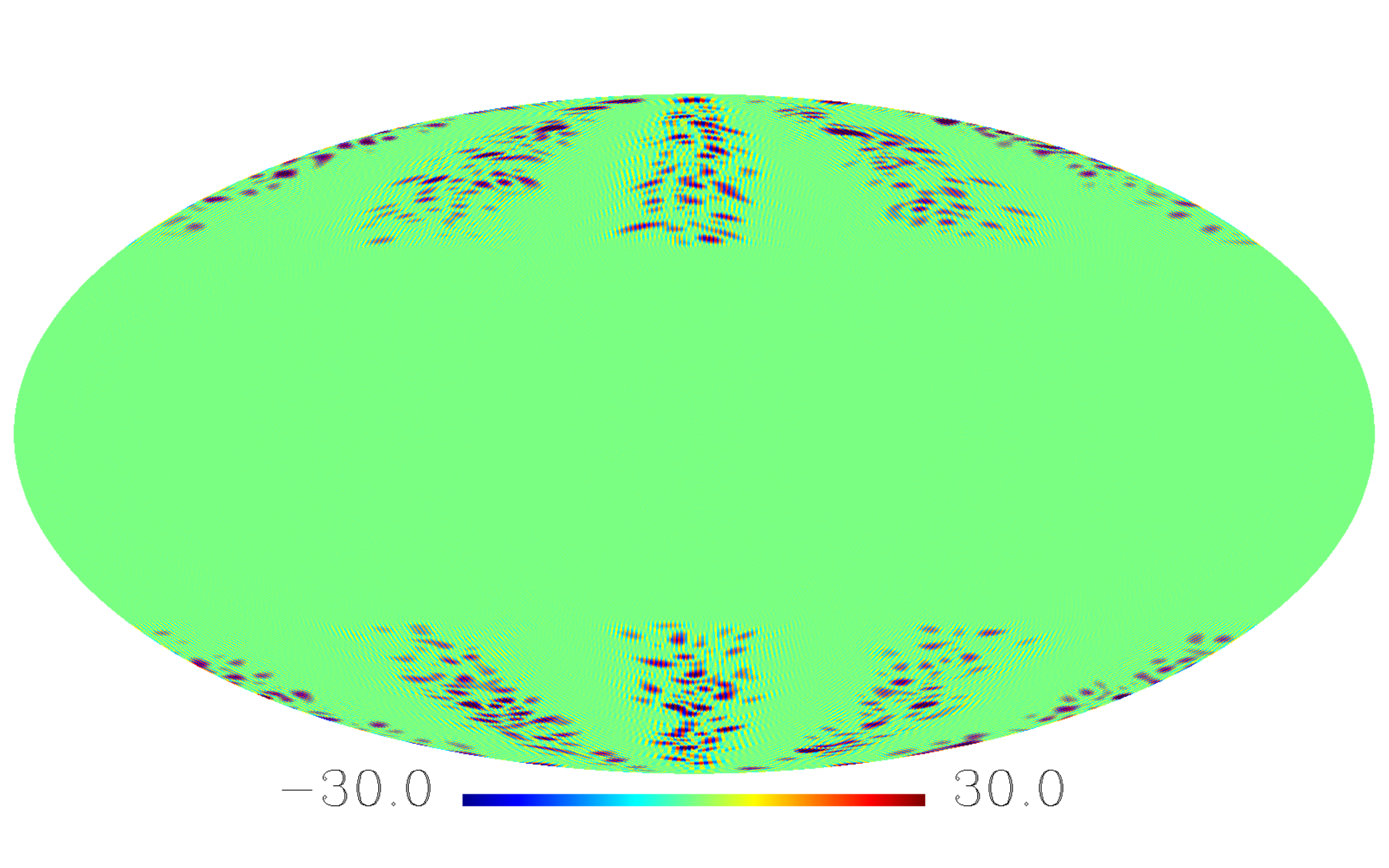}
  \includegraphics[width=0.23\textwidth]{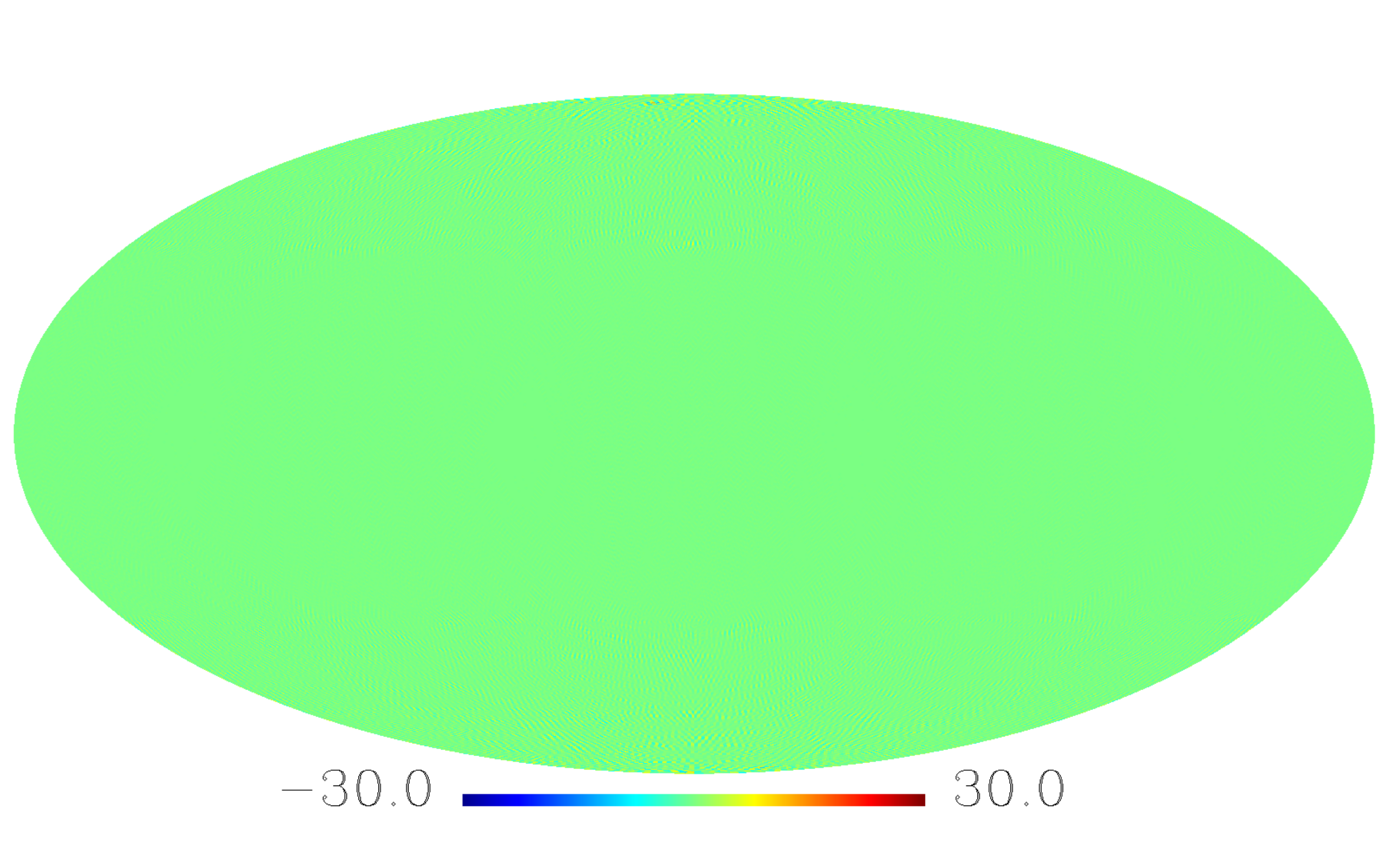}
  \caption{ The pixel domain errors of iterative solutions as $T_{\rm
  input}-T_{\rm output}$, for the traditional mode (left) and the
  immediate mode (right) with 3 rounds of iterations. The input maps are
  CMB intensity-only simulations based on the Planck 2018 best-fit CMB
  spectrum (\cite{2018arXiv180706209P}), with $N_{\rm side}=128$ a band
  limit of $\ell_{\rm max} \le 383$. }
  \label{fig:iter_error}
\end{figure}

We then compare these two iterative schemes in
Figure~\ref{fig:iter_error}, where the immediate mode is apparently more
accurate when the input maps are band-limited to $\ell_{\rm max} <
3N_{\rm side}$.\footnote{Note that if the input map has no band limit
(like a white noise map), the situation is more complicated and needs a
further study.} We also observe that the traditional mode gives better
convergence when the input maps have a stronger band-limit of $\ell_{\rm
max}\le 2N_{\rm side}$; however, in this case, the errors of both modes
are negligible.

\section{Summary and discussions}\label{sec:discussion}

In this work, we design a dedicated algorithm for SHTs on a large number
of sky maps, which is also a complete SHT toolkit for the CMB data,
including temperature, polarization; forward, backward; and iterative,
non-iterative transforms. The results of the new code are identical to
the traditional SHT toolkit like \textsl{Healpy}, but the performance is
significantly improved on a large number of sky maps. With the benchmark
environment described in section~\ref{sub:performance}: For the CPU
implementation, the speed of the new code is 2 $\sim$ 10 times of
\textsl{Healpy}; when GPU is employed, the typical acceleration factor
is more than one order of magnitude without iteration, and can be
further boosted to about 30$\times$ with 3 rounds of iterations. By
applying this new scheme to the standard fix-EB leakage pipeline for
2000 masked CMB maps at $N_{\rm side}=256$, even after including the
initialization overhead, the GPU version still gets a 10$\times$
speed-up compared to the \textsl{Healpy} SHT toolkit that does the same
job. Moreover, the acceleration factor of the GPU version increases
almost linearly with $N_{\rm side}$, indicating that the new code can be
much more efficient at a higher resolution (given enough GPU memory). As
far as we know, this is the fastest SHT implementation for a large
number of spherical maps till the present day.

However, we would like to point out the following facts: 1) A dedicated
compiler flag optimization can help to increase the performance of
\textsl{Healpy} by roughly 100\%; thus, in this case, the acceleration
factors should be divide by two. 2) The performance of \textsl{fastSHT}
is insensitive to the compiling flags, because the most time consuming
part (matrix multiplication) is done by either the Intel-MKL or NVIDIA
CUDA libraries, which are already optimized. 3) For a single input map,
the computation of SHT will degenerate from the BLAS level 3 to level 2,
so the core part of the new scheme will become less efficient. Moreover,
the acceleration requires using $P_{\ell m}$s as matrices, but for a
single map, it is more efficient to compute the $P_{\ell m}$s in small
segments to use the CPU L1 cache more efficiently. Similar reasons apply
for the GPU code, and the large initialization overheads also prevent
the \textsl{fastSHT-GPU} from outperforming \textsl{Healpy} when dealing
with only a small number of high-resolution maps. 4) Since it is
unlikely that one algorithm can be the best for all cases, and the
performances are significantly affected by the hardware and software
environments, we have provided a test script called ``run\_nproc.sh'' to
help readers run a batch of tests on their own machine to determine
which one is more suitable in a user-specified environment. 5) We are
also considering integrating the results of such test in the fastSHT to
make an automatic choice of the SHT engine for various computer hardware
and data scales.

We have also provided a new algorithm integrated in the \textsl{fastSHT}
to improve the accuracy of SHTs by optimizing the iteration strategy.
The test results show that this new iterative scheme is more accurate at
high multiples, especially for the temperature maps. We also point out
that when the input map is strictly band-limited to $\ell<2N_{\rm
side}$, the traditional iterative scheme is slightly better -- but both
iterative schemes nevertheless give negligible errors in this case.

Numerous ongoing and planned CMB experiments are going to bring heavy
data forecasting and analyzing tasks that usually involve processing
large amounts of mock data. As the core component in almost all CMB data
processing pipelines, the performance of SHTs has become critical. The
new scheme introduced in this paper fully exploits the state-of-the-art
software and hardware to accelerate SHTs. As a consequence, the time
costs for SHT operations are reduced considerably, and standard data
processing pipelines, such as fixing EB-leakages, can be accelerated
correspondingly. Also note that SHTs are widely used not only in
astrophysics and cosmology but in other areas of science as well,
including geoscience, atmospheric sciences, oceanography, etc., where
spherical data is used frequently. As a result, we anticipate that our
new scheme will have broader applications.

Finally, we would like to refer to the most recent development of the
multi-GPU computation technique that uses distributed GPU memory in a 2D
block-cyclic fashion~\citep{cuBLASNV74:online}. This technique is still
in an early view stage, and it is expected to provide access to much
more GPU memory and computational power in a dedicated way for matrix
multiplications; hence it will likely solve the GPU memory size problem
that prevents our \textsl{fastSHT-GPU} code from working sufficiently
well at a very high resolution, like $N_{\rm side}\ge 2048$. We shall
keep tracking the development of this technique and update our code
accordingly in the future.

\begin{acknowledgments}

This work is supported by the National Key R\&D Program of China
(2021YFC2203100, 2021YFC2203104) and the Anhui project Z010118169. We
would like to thank the referee's professional comments that helped
improve this work. We also thank Matthew Carney, James Mertens, and
Glenn Starkman for helpful discussions, and our USTC colleagues for
helping us with the GPU test environment.

\end{acknowledgments}

\appendix

\section{A few technical tips}\label{app:tech tips}

\subsection{FFT-mapping with the unit circle}
\label{appsub:fft-mapping with unit circle}

\begin{figure}[!htb]
  \centering
  \includegraphics[width=0.22\textwidth]{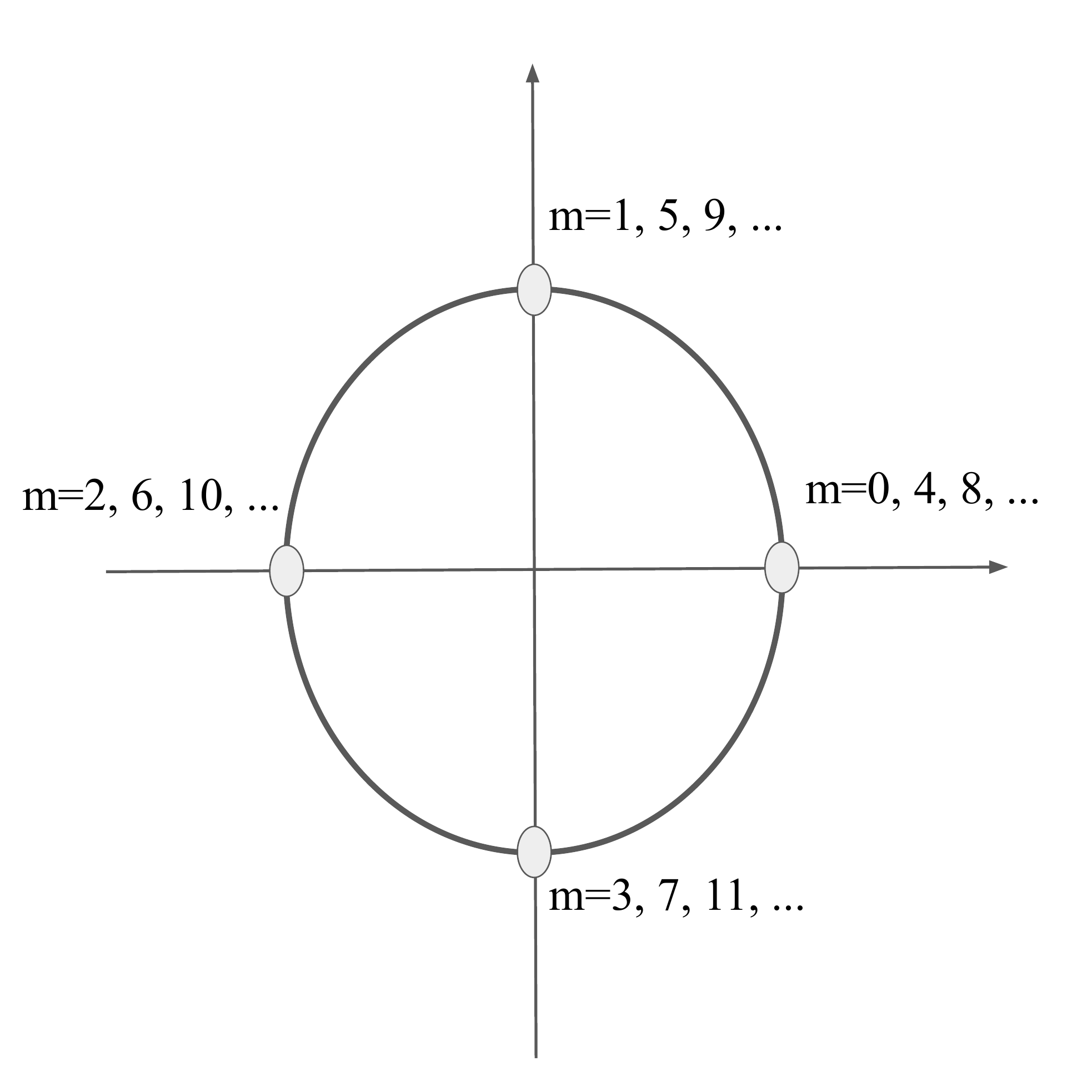}
  \includegraphics[width=0.22\textwidth]{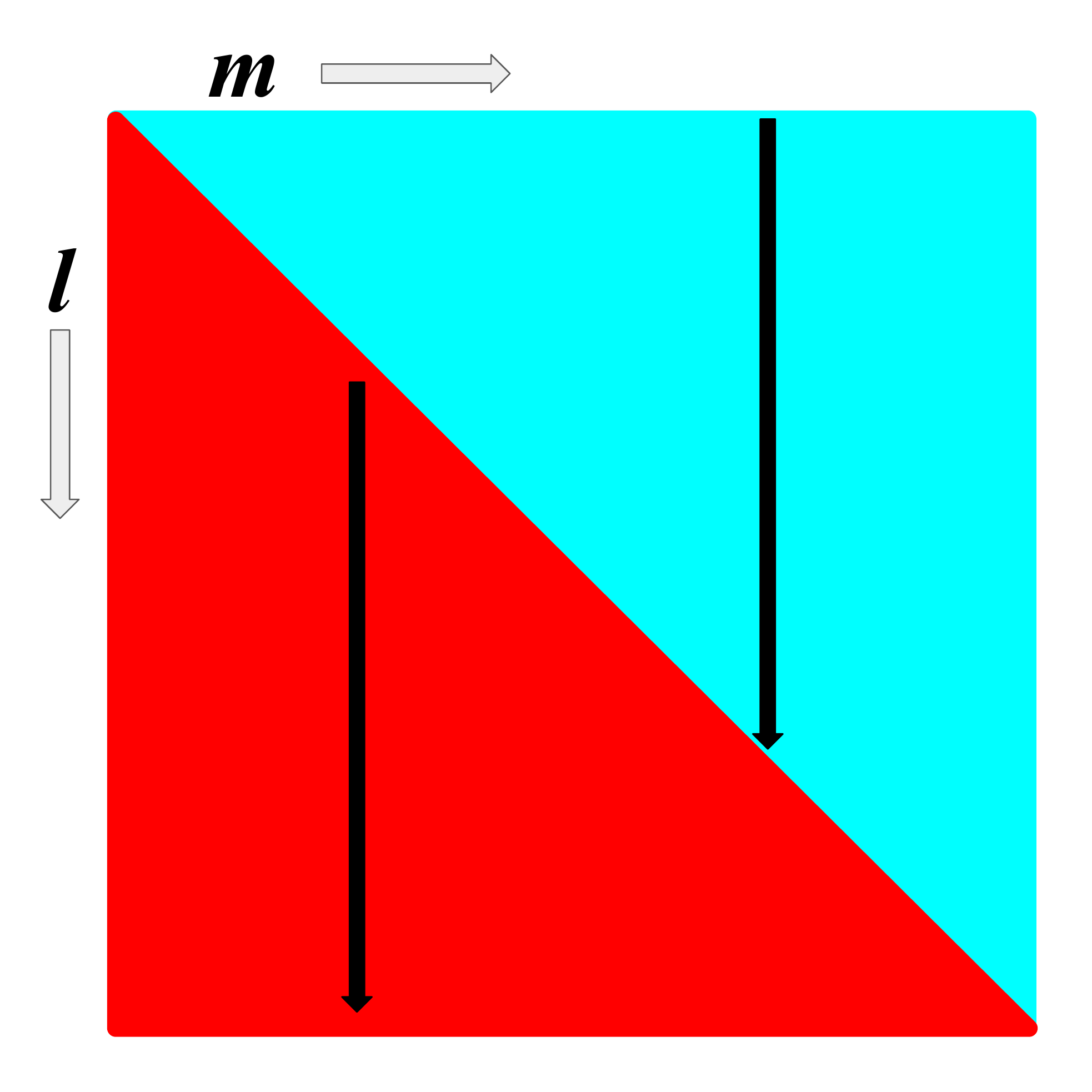}
  
  \caption{ \emph{Left}: An example of the FFT-frequency circle showing
  the mapping of different $m$ when the FFT size is $N=4$. \emph{Right}:
  Illustration of the new saving scheme of $a_{\ell m}$, adopted to
  match the requirement of a large number of matrix operations. The
  lower half (red) is for the real part, and the upper half (cyan) is
  for the imaginary part. The real and imaginary parts of $a_{\ell m}$
  for one $m$ are distributed as shown by the black arrows. }
  \label{fig:fft-circle}
\end{figure}

One thing to be noticed in a SHT is that, in most cases, the number of
pixels in one ring and the value of $m$ do not match, thus one should
properly map $m$ to the FFT frequency circle, as shown in
Figure~\ref{fig:fft-circle}, which is a unit circle centered around the
original point, with points on it marked by the azimuthal angles $\phi$.
In the frequency circle, $\phi=0$ correspond to the zero frequency,
$\phi=\pi$ correspond to the Nyquist frequency, and all frequencies
above the Nyquist frequency will be mapped to lower frequencies by
moving along the circle in an anti-clockwise direction. This is also a
clear presentation of the Nyquist-Shannon sampling theorem, which
naturally shows why one cannot exceed the Nyquist frequency. As an
example, $m=0$ is always mapped to $\phi=0$, regardless of the value of
$N$ (size of one ring); but other $m$ should be mapped to
$\phi=\frac{2\pi m}{N}$. This means, even if $m$ is the same, in rings
with different $N$, it will be mapped to different positions, and vice
versa.

\subsection{ Polynomials for polarization }
\label{appsub:pol plm}

In the context of rotation, $Q$ and $U$ can be decomposed into spin
$\pm2$ spherical harmonics~\cite{PhysRevD.55.1830, 0004-637X-503-1-1} as
follows:
\begin{equation} \label{Q_lm+iU_lm 1}
    Q(\mathbf{\hat{n}})\pm i U(\mathbf{\hat{n}}) = 
    \sum_{l,m} a_{\pm2,lm}\;{}_{\pm2}Y_{lm}(\mathbf{\hat n}),
\end{equation}
where $_{\pm2}Y_{lm}(\mathbf{\hat n})$ are the spin $\pm2$ weighted
spherical harmonics. Thus, the spin $\pm2$ spherical harmonic
coefficients $a_{\pm2,lm}$ are given by the following:
\begin{align}
a^k_{\pm2,lm} = \sum_{\theta}P_{\pm2,\ell m}(\cos(\theta))
(\mathcal{Q}^k_{im}\pm i\,\mathcal{U}^k_{im}),
\end{align}
where $\mathcal{Q}^k_{im}$ and $\mathcal{U}^k_{im}$ are the Fourier
transforms of the $Q$ and $U$ Stokes parameters of the $k$-th map at the
$i$-th ring. For simplicity, we define the following matrix forms:
\begin{align}
\mathbf{A}^\pm_{m} = (a^k_{\pm 2,\ell})_m
,\;
\mathbf{P}^\pm_{m} = (P_{\pm 2,\ell})_m
\end{align}

Because the $E$- and $B$-mode spherical harmonic coefficients are
defined as
\begin{align} \label{equ:alm-eb 1} 
a_{E,lm} &= -(a_{2,lm} + a_{-2,lm})/2,
\nonumber \\ 
a_{B,lm} &= i(a_{2,lm} - a_{-2,lm})/2,
\end{align}

In this work, we use a \textsl{HEALPix}-like program to compute the
$P_{\ell m}$ coefficients for both temperature and polarization. One
should note that (seems missing in the \textsl{HEALPix}  manual), the
convention is to compute the summation and difference of the polarized
$P_{\ell m}$ coefficients, which are used in SHTs instead of polarized
$P_{\ell m}$s themselves. With such convention, the polarized $a_{\ell
m}$s are computed from the Q and U stokes parameters by
\begin{align}\label{equ:healpix pol convension}
-F^{+}_{\ell m}(\sigma) &= \frac{1}{2}\left[Y_{+2,\ell m}(\sigma) + 
Y_{-2,\ell m}(\sigma)\right] = \lambda_{\ell m}^{+}(\theta)e^{\bm{i}m\phi} \\ \nonumber
-F^{-}_{\ell m}(\sigma) &= \frac{}{2}\left[Y_{+2,\ell m}(\sigma) - 
Y_{-2,\ell m}(\sigma)\right] = \lambda_{\ell m}^{-}(\theta)e^{\bm{i}m\phi},
\end{align}
where $Y_{\pm2,\ell m}$ are the spin$\pm2$ spherical harmonics, and
$\lambda_{\ell m}^{\pm}(\theta)$ are the effective Legendre coefficients
generated by \textsl{HEALPix} . Let $X$ be either $Q$ or $U$, then the E
and B-mode spherical harmonic coefficients, computed with the
\textsl{HEALPix}  convention, are
\begin{align}\label{equ:healpix pol convension2}
X_{\ell m}^{\pm} &= \int X(\sigma) F^{\pm*}_{\ell m}(\sigma) d\sigma\\ \nonumber
a_{\ell m}^E &= -Q_{\ell m}^{+} - \bm{i}U_{\ell m}^{-} \\ \nonumber
a_{\ell m}^B &= -U_{\ell m}^{+} - \bm{i}Q_{\ell m}^{-} .
\end{align}

Lastly, a new storage scheme of $a_{\ell m}$ is introduced to match the
requirement of high performance computing with many input maps, in which
all $a_{\ell m}$ coefficients are stored in a real-valued square matrix
of size $(\ell_{\textrm{max}}+1)\times(\ell_{\textrm{max}}+1)$. The real
and imaginary parts are compactly\footnote{Here ``compact'' means all
matrix elements are used without redundancy.} stored in the square
matrix with a very simple addressing rule (see also
Figure~\ref{fig:fft-circle}):
\begin{align}\label{equ:new alm save scheme}
\begin{matrix*}
\textrm{Real part:}      & \,\,(\ell,m)\,\, \Longrightarrow& (\ell,m)  \\
\textrm{Imaginary part:} & \,\,(\ell,m)\,\, \Longrightarrow& (\ell-m, \ell_{\textrm{max}}-m+1), \\
\end{matrix*}
\end{align}
where we have the natural constraints that all non-zero imaginary parts
satisfy: $1 \leqslant m \leqslant\ell_{\textrm{max}}$ and $m \leqslant
\ell \leqslant\ell_{\textrm{max}}$. This new scheme is especially
important for the GPU-implementation, because this kind of regular shape
tensor-style storage with very simple addressing rule is apparently
favored by the GPU. Even in a CPU-implementation, it is able to
significantly reduce the time cost of memory operations.

\section{The possibility of allowing more iterations}

We know that the accuracy of SHTs can be improved with an increasing
number of iteration rounds. The default choice in \textsl{Healpy} is 3
rounds, which is a balance between the time cost and accuracy -- with
the ``time cost'' estimated under the old
scheme~\citep{map2almi42:online}. However, \textsl{fastSHT-GPU} is
significantly more efficient with a bigger number of iterations, because
the iterative part of SHT is done in GPU instead of CPU. This fact will
break the old balance and allow more iterations at a reasonable time
cost. In order to evaluate this effect, We have done a non-polarization
\textsl{fastSHT-GPU} test with $N_{\rm side}=256$, $\ell_{\rm
max}=3N_{\rm side}-1$, $N_{\rm maps}=1000$ and $N_{\rm iter}=3$ and $9$
respectively: When $N_{\rm iter}=3$, the time cost of the non-iterative
part (the operations outside the iteration) is and each round of
iteration (including two SHTs, one forward and one backward) costs only
$0.444\, s$. Thus we can predict the time cost with any number of
iterations by the following equation:
\begin{align}
    t_{\rm total} = t_0 + (2N_{\rm iter}+1)t_1,
\end{align} 
and for the example here, we have $t_0=0.592\, s$ and $t_1=0.222\,s$
respectively. To double check the above equation: for $N_{\rm iter}=9$,
the total time cost is $4.800\,s$, which is consistent with the time
computed from the above equation. Therefore, when $N_{\rm iter}$
increase from 3 to 9, the time cost only increases from $2.1\,s$ to
$4.8\,s$, indicating that we get a relatively higher efficiency when
$N_{\rm iter}$ increases for \textsl{fastSHT-GPU}.

\bibliography{manual_ref}

\end{document}